\documentclass[12pt]{paper}
\usepackage[margin=0.9in]{geometry}
\usepackage{amsmath}
\usepackage{luatex85}
\usepackage{hyperref}
\usepackage{bm}
\usepackage{hhline}
\usepackage{xfrac}
\usepackage{caption}
\usepackage{subcaption}
\usepackage{amssymb}
\usepackage{mathtools}
\usepackage{tcolorbox}
\usepackage[bitstream-charter]{mathdesign}
\DeclareMathSymbol{\shortminus}{\mathbin}{AMSa}{"39}

\usepackage{titletoc,tocloft}
\dottedcontents{section}[1.5em]{}{1.3em}{.6em}

\title{Studies of multi-jet merging with Parton Branching TMD evolution}
\subtitle{Presented at DIS2022: XXIX International Workshop on Deep-Inelastic Scattering and Related Subjects, Santiago de Compostela, Spain, May 2-6 2022}
\author{Aron Mees van Kampen (University of Antwerp)}
\makeatletter\chardef\pdf@shellescape=\@ne\makeatother

\DeclareSymbolFont{usualmathcal}{OMS}{cmsy}{m}{n}
\DeclareSymbolFontAlphabet{\mathcal}{usualmathcal}

\usepackage{cite,./mcite}

\begin{document}

\maketitle

% For convenience during refereeing (optional),
% you can turn on line numbers by uncommenting the next line:
%\linenumbers

%\definecolor{palegray}{gray}{0.95}
%\begin{center}
%\colorbox{palegray}{
%  \begin{tabular}{rr}
%  \begin{minipage}{0.1\textwidth}
%    \includegraphics[width=22mm]{Logo}
%  \end{minipage}
%  &
%  \begin{minipage}{0.75\textwidth}
%    \begin{center}
%    {\it Proceedings for the XXIX International Workshop\\ on Deep-Inelastic Scattering and
%Related Subjects,}\\
%    {\it Santiago de Compostela, 2-6 May 2022} \\
%    \doi{}\\
%    \end{center}
%  \end{minipage}
%\end{tabular}
%}
%\end{center}

\section*{Abstract}
{\bf
QCD predictions for final states with multiple jets in hadron collisions 
make use of multi-jet merging methods, which allow one to combine 
consistently the contributions from hard scattering matrix elements with different parton multiplicities and 
parton showers. In this article I consider a multi-jet merging method that 
has recently been proposed to take into account the effects of transverse 
momentum dependent (TMD) evolution and parton shower, and present 
studies focusing on the application of this method to jets associated with 
Drell-Yan (DY) production in the region of high masses. 
}

%\vspace{10pt}
%\noindent\rule{\textwidth}{1pt}
%\tableofcontents\thispagestyle{fancy}
%\noindent\rule{\textwidth}{1pt}
%\vspace{10pt}

\section{Introduction}
\label{sec:intro}
%Increasing precision and accuracy of high energy hadron collisions require improvement in many areas. 
%One could categorize these into two main categories: QCD theory and numerical algorithms. 
%In this article, I concentrate on aspects within both categories: transverse momentum dependent (TMD) factorization as part of QCD theory and multi-jet merging as a numerical tool to combine different parts of the theory. 

Drell-Yan (DY) plus jets final states are important at the LHC for tests of the standard model as well as for backgrounds to Higgs studies and beyond standard model studies.  Baseline predictions for this are obtained by perturbative fixed-order calculations combined with parton showers in Monte Carlo event generators. Such predictions employ PDFs, which describe initial-state partonic structure and evolution within the collinear approximation.

In certain regions of the DY plus jets phase space, however, it becomes important to treat initial state QCD effects by including transverse momentum dependent (TMD) distributions \cite{Angeles-Martinez:2015sea}, going beyond collinear PDFs. TMD effects in DY plus jets have been the subject of recent studies \cite{Martinez:2022wrf, Chien:2022wiq, Yang:2022qgk, Buonocore:2022mle, Martinez:2021chk, Chien:2019gyf, Sun:2018icb, Buffing:2018ggv}.

In this article I apply the TMD jet merging method \cite{Martinez:2021chk} to study the differential jet rates (DJRs) associated with DY production as a function of the DY mass. In particular, I consider the region of high DY masses, and investigate the behavior of the TMD merging method and the merging scale in this new mass region.

%The combination of matrix elements at high orders in the strong coupling $\alpha_s$ and soft contributions from e.g. parton showers is referred to as matching and merging. These higher orders are needed to describe events that include (multiple) jets. 

This article is organized as follows. I introduce the theoretical framework for TMD evolution and TMD merging in section \ref{sec:theory}. In section \ref{sec:sliding} I present numerical calculations of the DJRs and discuss their main features. In section \ref{sec:conclusion} I give conclusions.

\section{Parton Branching method and TMD merging} \label{sec:theory}
The parton branching (PB) method developed in \cite{Hautmann:2017xtx, Hautmann:2017fcj} describes the evolution of TMDs by means of real-emission splitting functions, Sudakov form factors and angular ordering phase space constraints. Nonperturbative distributions at the initial scale of order 1 GeV are determined from fits to experimental data \cite{BermudezMartinez:2018fsv}. For applications, the evolved TMD distributions are matched with fixed-order hard-scattering matrix elements \cite{Yang:2022qgk,Martinez:2019mwt}. The TMD distributions and corresponding TMD parton shower are implemented in the Monte Carlo event generator \textsc{Cascade3} \cite{Baranov:2021uol}.

Applications to DY transverse momentum spectra, using PB TMDs matched with next-to-leading-order (NLO) DY matrix elements via MC@NLO \cite{Alwall:2014hca}, have been studied in\cite{Martinez:2019mwt}  for LHC energies and in \cite{BermudezMartinez:2020tys} for lower energies. Applications for jets have been studied in \cite{Abdulhamid:2021xtt} by the same method.

It is found in \cite{Martinez:2019mwt} that NLO-matched PB TMD predictions provide a very good description of experimental LHC measurements of the Z-boson transverse momentum $q_T$ distribution in the low $q_T$ region and middle $q_T$ region ($q_T \lesssim M_Z$) while a deficit is observed in the prediction compared to experimental data in the high $q_T$ region. The low $q_T$ region is dominated by soft-gluon radiation emitted through TMD evolution. The high $q_T$ region is dominated by hard, perturbative emissions, and the deficit points to the lack of higher orders beyond the first hard emission in an NLO calculation. 

A method to include higher order emissions corresponding to Z + n partons has been devised in \cite{Martinez:2021chk} based on multi-jet merging. The method extends the MLM merging\cite{Mangano:2006rw, Alwall:2007fs, Mrenna:2003if, Mangano:2002bhl} to the case of TMD parton branching, rather than collinear parton branching. It allows one to combine consistently, without double counting or missing events, high multiplicity matrix elements with the TMD parton showers and TMD parton distributions. Using the TMD merging, a good description of the Z-boson $q_T$ spectrum is obtained not only at low $q_T$ but also at high $q_T$. This observation is also made by the recent CMS collaboration analysis \cite{CMS:2022ilp}.

The key features of the TMD jet merging, compared to the collinear MLM merging are that, owing to taking into account the transverse momentum in the initial state parton cascade: i) a better description of high multiplicity jet final states is achieved, and ii) a reduction of the systematic uncertainties due to the merging parameters is observed in the theoretical predictions. 

In refs. \cite{Martinez:2021chk,Martinez:2021dwx}, multi-jet observables in Z + jets production are studied using the TMD merging in the region near the Z boson mass. 
In the next section, calculations with TMD merging away from the Z-boson mass are shown to explore the behavior of the TMD merging approach and its merging scale in the region of high masses.  

%Observables as the Drell-Yan $p_T$ spectrum and azimuthal decorrelation in dijet events are also sensitive to soft gluon emissions and the perturbative series need to be resummed to all orders in the strong coupling. 
%
%The parton branching method \cite{Hautmann:2017xtx, Hautmann:2017fcj, BermudezMartinez:2018fsv, Hautmann:2019biw, BermudezMartinez:2019anj, BermudezMartinez:2020tys} provides an approach for the evolution of TMD PDFs in a Monte Carlo manner. This lies at the basis to implement TMD factorization in event generators and has been implemented in \textsc{CASCADE3} \cite{Baranov:2021uol}. This formalism includes resummation of soft-gluon emissions up to NLL accuracy \cite{vanKampen:2021oxe}. 
%After the achievement of matching PB to NLO matrix elements \cite{BermudezMartinez:2019anj}, recently a method for multi-jet merging with PB TMDs has been developed \cite{Martinez:2021chk}. 

\section{DJRs in DY + jets at high masses}
\label{sec:sliding}
Differential jet rates (DJRs) are the distributions of the variable $d_{n,n+1}$, which is the square of the energy scale at which an $n$-jet configuration is resolved as an $(n+1)$-jet configuration, with jets defined according to the kT jet algorithm \cite{CATANI1993187,Ellis:1993tq}. They are sensitive to the consistency of any jet merging method, which makes them appropriate quantities to validate the merging algorithm.

The calculational set-up is as follows. With MadGraph5\_aMC@NLO \cite{Alwall:2014hca} I generate Les Houches Event (LHE) files at leading order (LO) containing hard scattering events - that represent the matrix elements - for pp collisions to Z + 0, 1, 2, 3 jet final states at a centre-of-mass energy of $\sqrt{s} = 13$ TeV. A generation cut of 16 GeV sets the lower limit of emitted transverse momentum by the partons in these events. The transverse momentum at the hard scale is generated by forward evolution \cite{Hautmann:2017xtx} using the PB-TMD Set 2 \cite{BermudezMartinez:2018fsv} provided by the TMDlib library \cite{Abdulov:2021ivr, Hautmann:2014kza}. Parton shower emissions are generated in the TMD shower in \textsc{Cascade} \cite{Baranov:2021uol} following the PB evolution dynamics in a backwards manner. The TMD merging algorithm \cite{Martinez:2021chk} at LO is implemented within the event generator. Hadronization is turned off, since this enters the calculation only after the merging and would not affect the DJRs.

The merging scale in a merged calculation (indicated with $E_\perp^{\text{clus}}$) represents the minimal transverse energy of a jet to pass the merging algorithm. The effect on the DJR is that the region below the merging scale is dominated by the TMD and TMD shower, while the region above the merging scale is dominated by the matrix element. 

\begin{figure}[h!]
    \centering
    \begin{subfigure}[b]{0.32\textwidth}
         \includegraphics[width=0.95\textwidth]{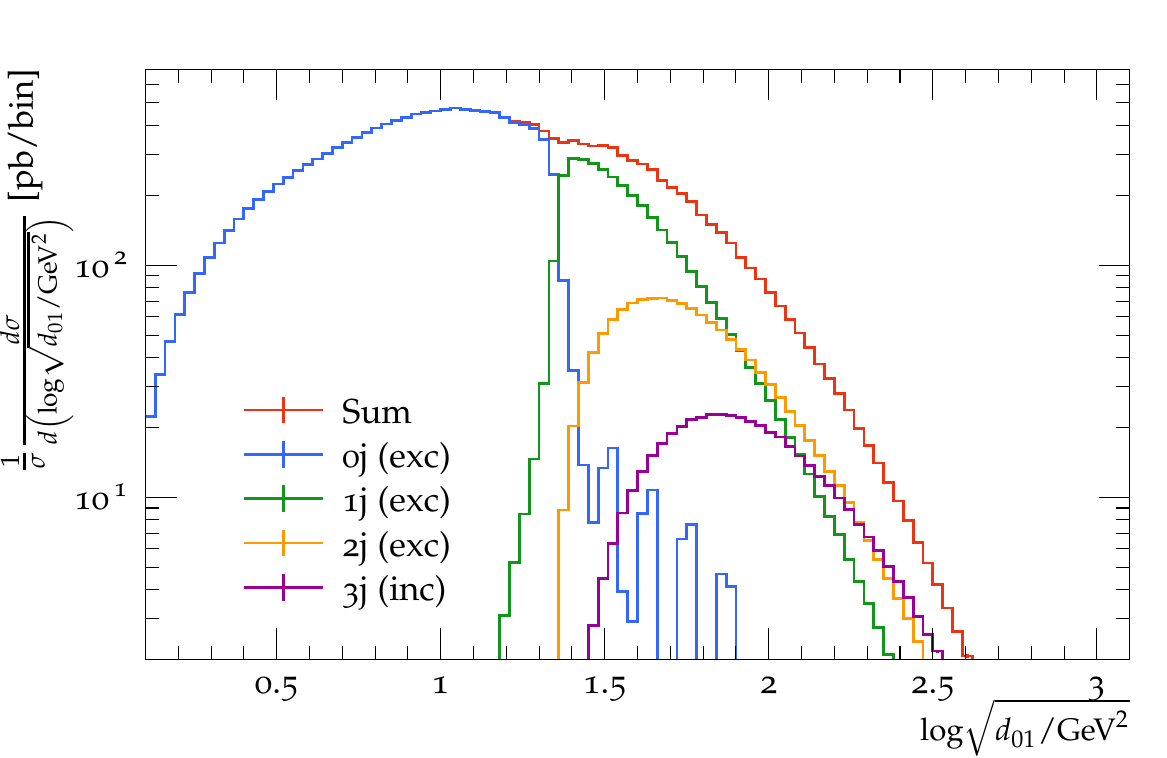}
    \label{DJR01-MZ_etclus23}
    \caption{$d_{01}$}
    \end{subfigure}
     \begin{subfigure}[b]{0.32\textwidth}
         \centering
        \includegraphics[width=0.95\textwidth]{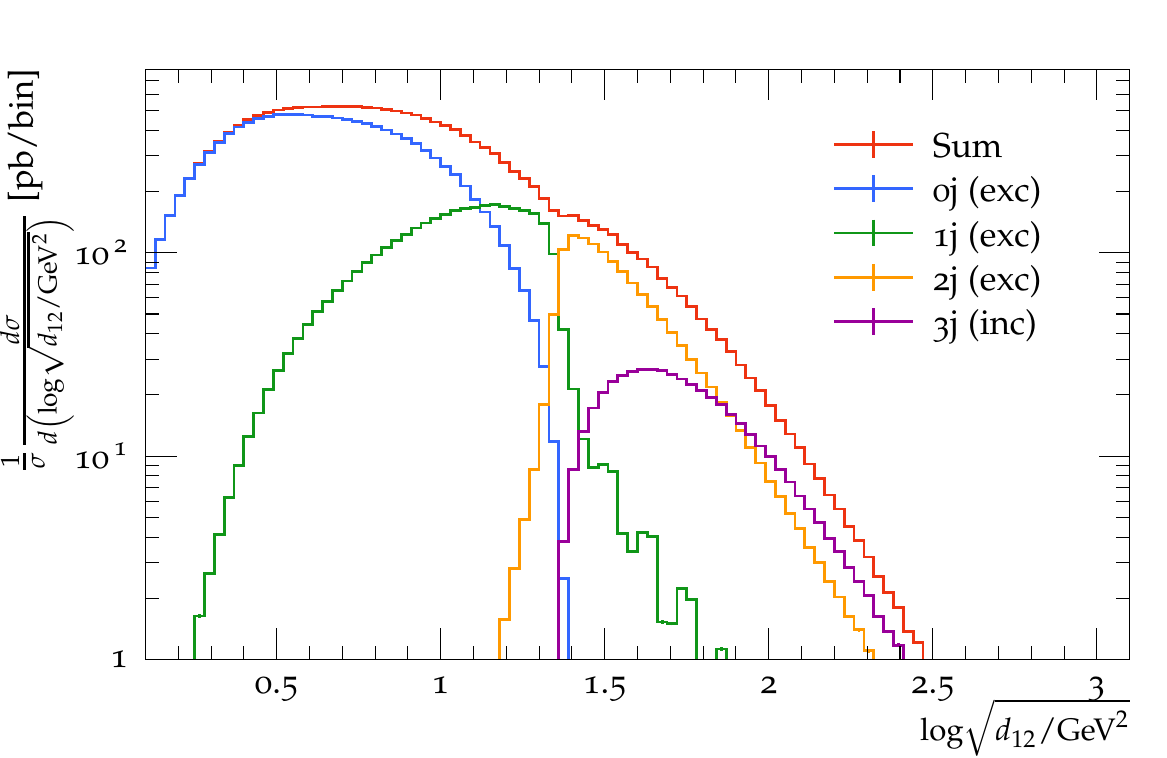}
            \label{DJR12-MZ_etclus23}
            \caption{$d_{12}$}
     \end{subfigure}
     \begin{subfigure}[b]{0.32\textwidth}
         \centering
         \includegraphics[width=0.95\textwidth]{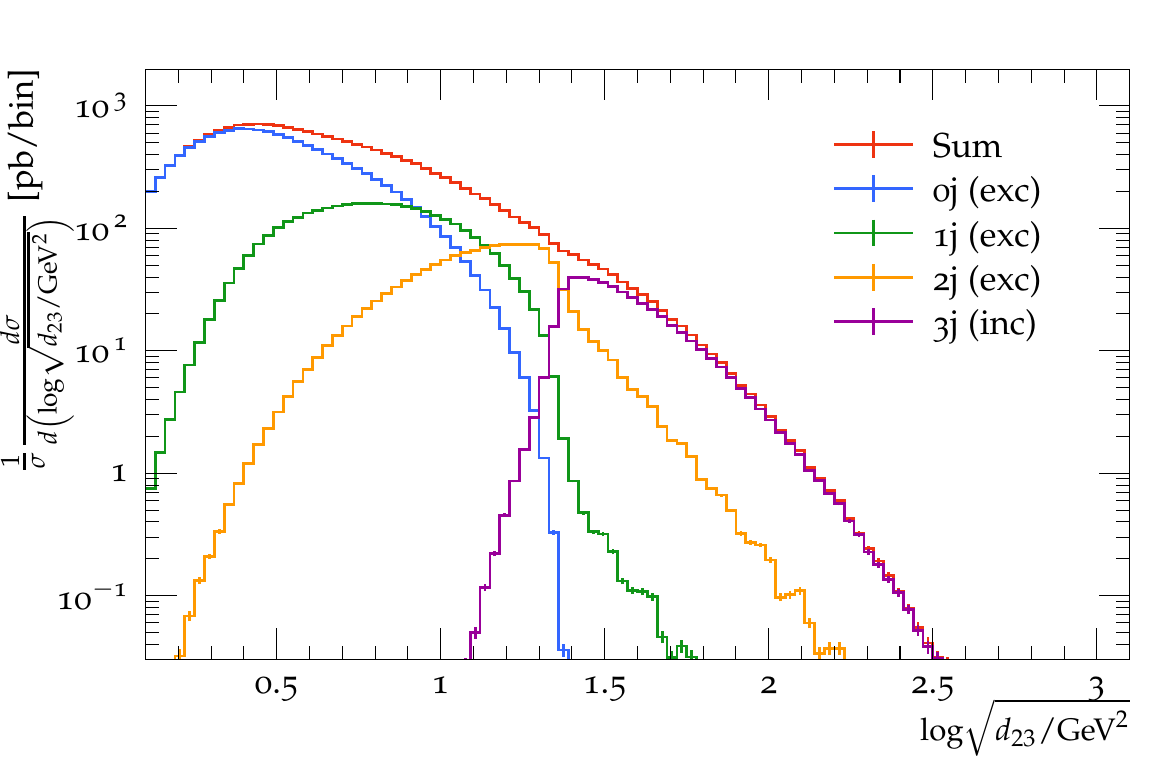}
     \label{DJR23-MZ_etclus23}
     \caption{$d_{23}$}
     \end{subfigure}
     \caption{DJRs for Z+jets around the Z boson mass with a merging scale $E_\perp^{\text{clus}}=23$ GeV.}
     \label{fig:MZetclus23}
\end{figure}

We investigate DJRs at different hard interaction scales to study the behavior of the merging scale when the hard scale varies. 
%In case no restrictions are applied to the hard scale, 
For di-lepton mass around the $Z$-boson mass, 
a merging scale of 23 GeV has been applied~\cite{Martinez:2021chk}. 
This    works  well for TMD merging of Z+jets as observed in \cite{Martinez:2021dwx}. Figure \ref{fig:MZetclus23} shows that the DJRs are smooth in this regime. 
%There is no lower limit on the di-lepton mass, which means that the mass distribution peaks at $M_Z\simeq 91.2$ GeV.

\begin{figure}[h!]
    \centering
    \begin{subfigure}[b]{0.32\textwidth}
         \includegraphics[width=0.95\textwidth]{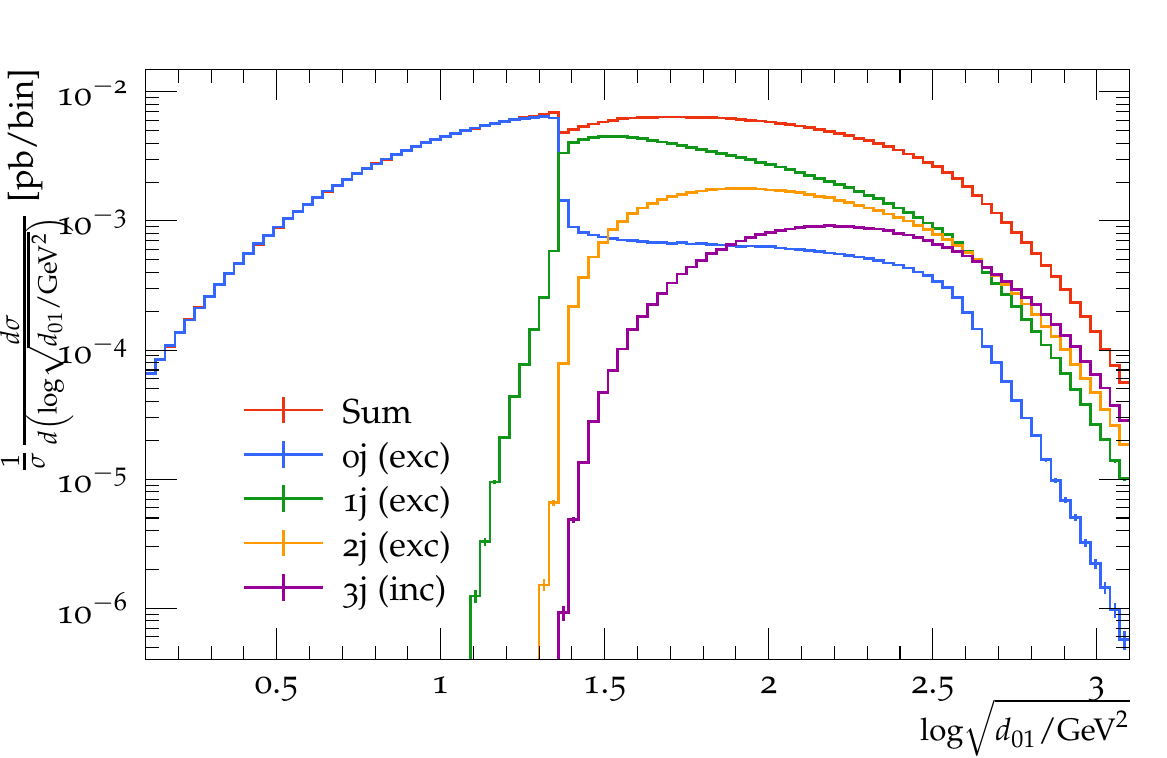}
    \label{DJR01_mmll800_etclus23}
    \end{subfigure}
     \begin{subfigure}[b]{0.32\textwidth}
         \centering
        \includegraphics[width=0.95\textwidth]{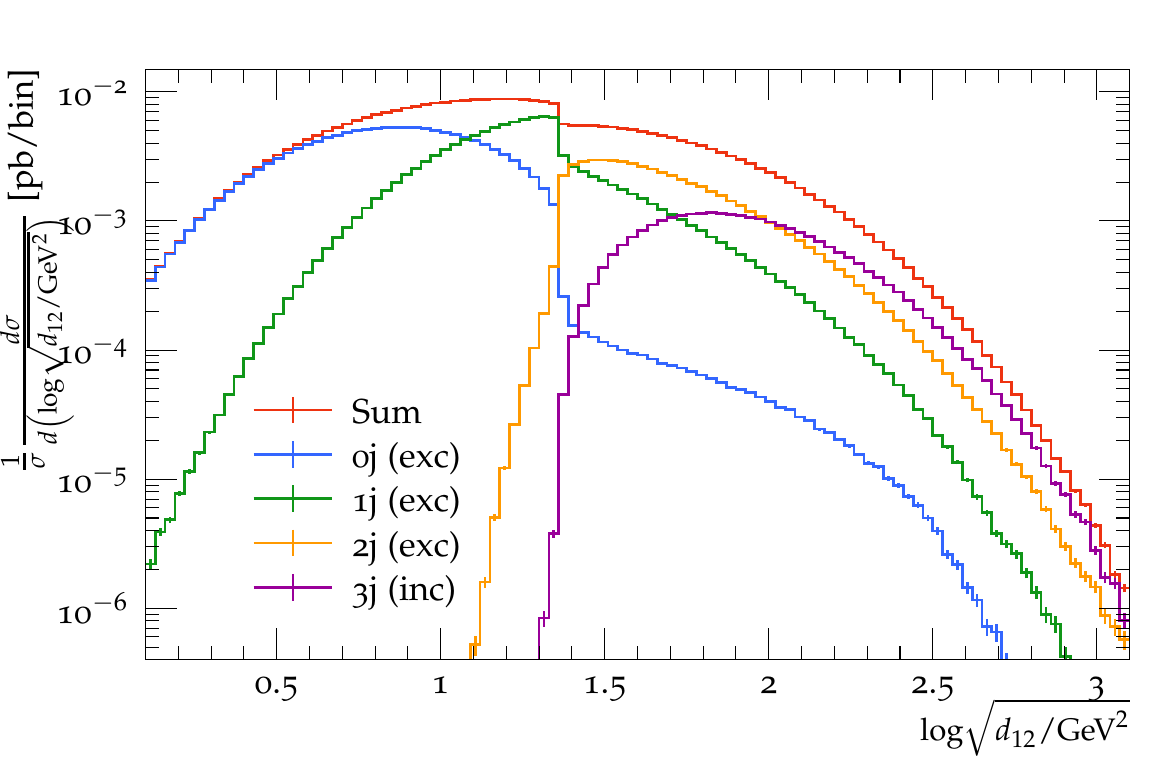}
            \label{DJR12_mmll800_etclus23}
     \end{subfigure}
     \begin{subfigure}[b]{0.32\textwidth}
         \centering
         \includegraphics[width=0.95\textwidth]{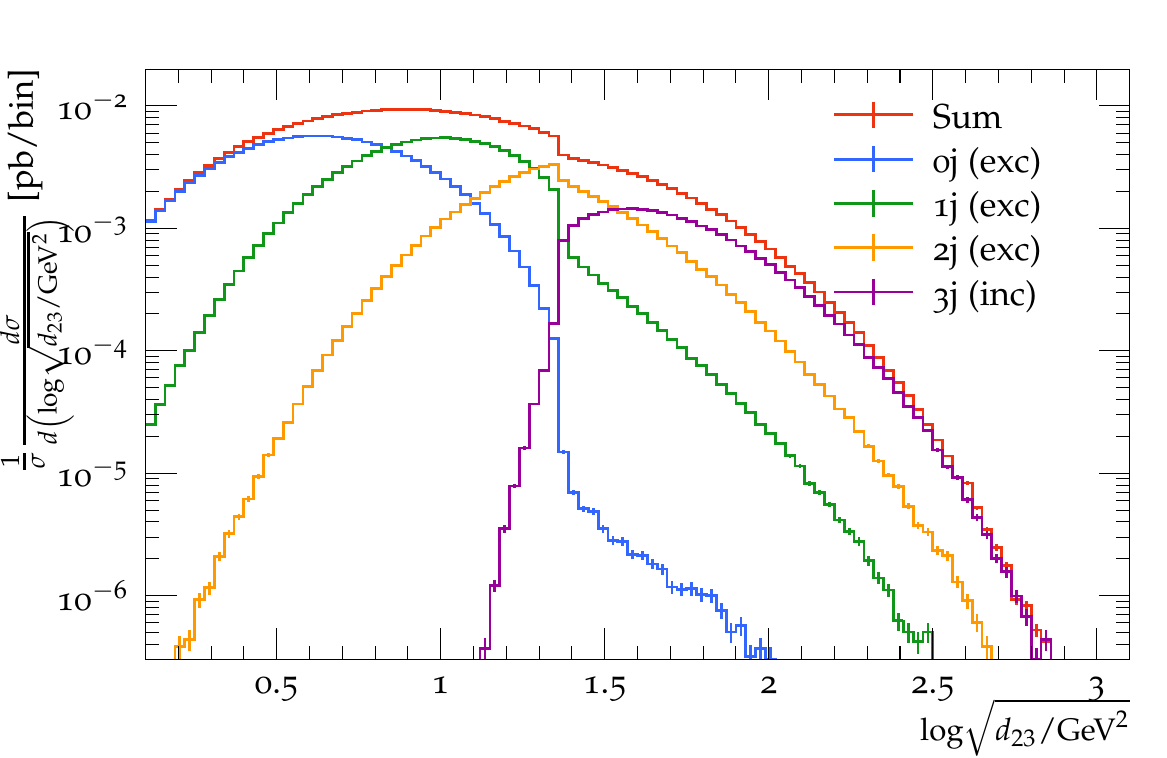}
     \label{DJR23_mmll800_etclus23}
     \end{subfigure}
     \caption{DJRs for a minimal DY mass $M_{ll}^{\text{min}}=800$ GeV and a merging scale $E_\perp^{\text{clus}}=23$ GeV.}
     \label{fig:etclus23}
\end{figure}

\begin{figure}[h!]
    \centering
    \begin{subfigure}[b]{0.32\textwidth}
         \includegraphics[width=0.95\textwidth]{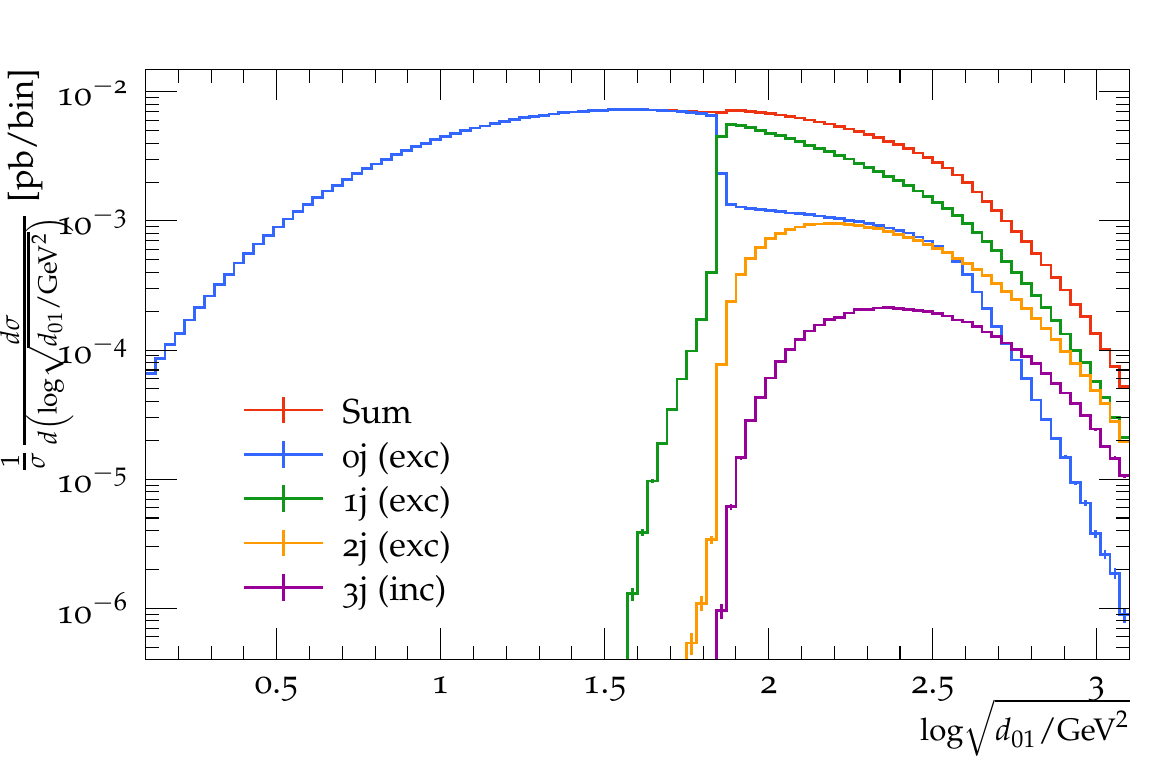}
    \label{DJR01_mmll800_etclus70}
    \end{subfigure}
     \begin{subfigure}[b]{0.32\textwidth}
         \centering
        \includegraphics[width=0.95\textwidth]{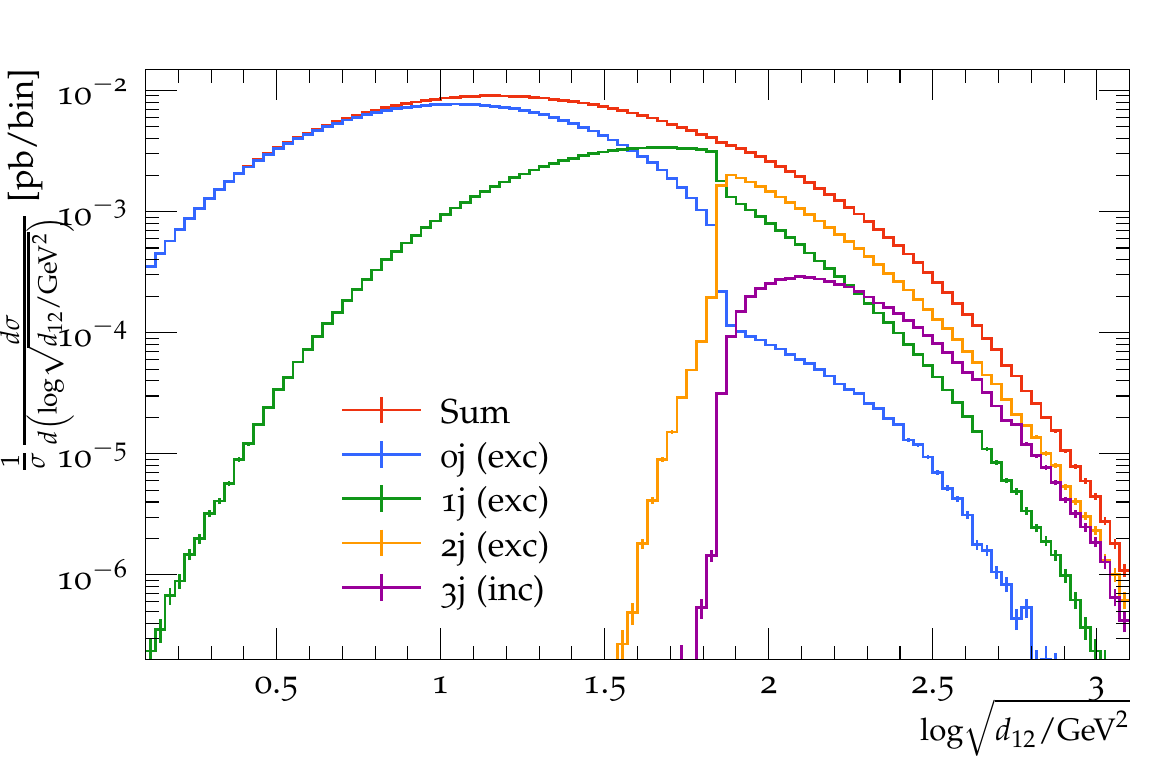}
            \label{DJR12_mmll800_etclus70}
     \end{subfigure}
     \begin{subfigure}[b]{0.32\textwidth}
         \centering
         \includegraphics[width=0.95\textwidth]{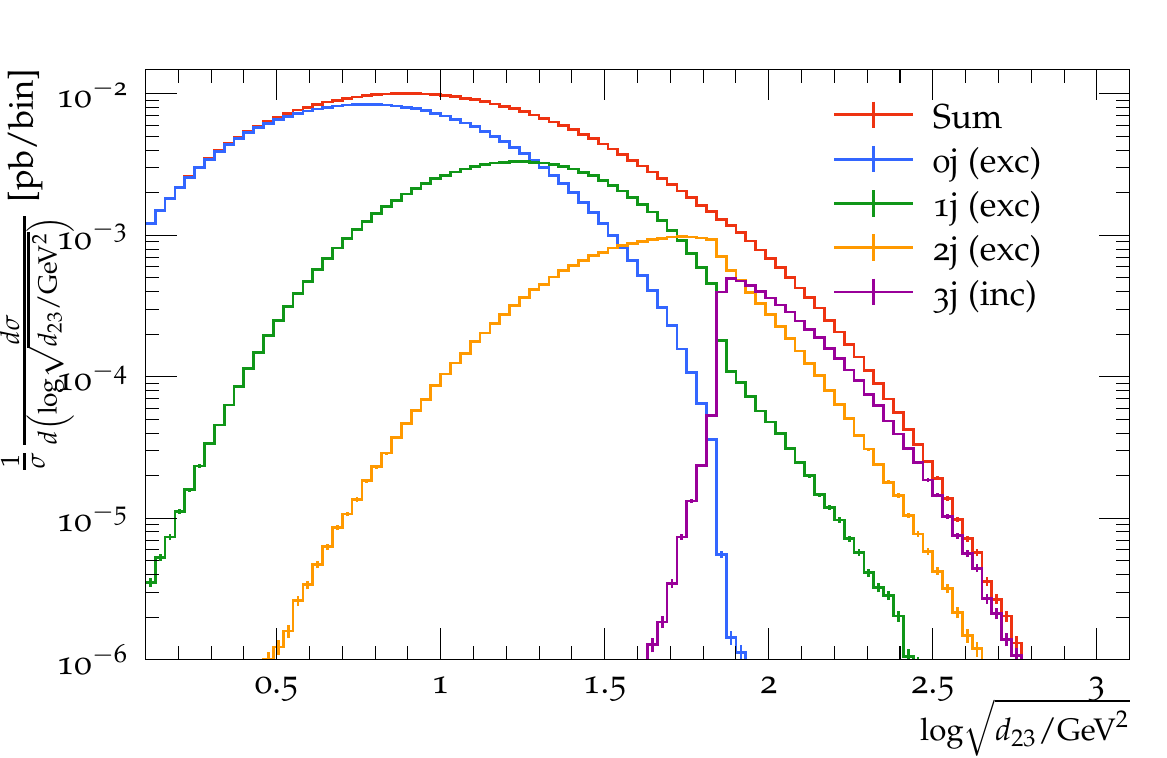}
     \label{DJR23_mmll800_etclus70}
     \end{subfigure}
     \caption{DJRs for a minimal DY mass $M_{ll}^{\text{min}}=800$ GeV and a merging scale $E_\perp^{\text{clus}}=70$ GeV.}
     \label{fig:etclus70}
\end{figure}

In Figures \ref{fig:etclus23} and \ref{fig:etclus70}, we show the DJRs resulting from studies in the scenario where the minimal mass of the di-lepton pair is set to 800 GeV. At the merging scale  of 23 GeV ($\log(23)=1.36$), a discontinuity occurs (Fig. \ref{fig:etclus23}). To resolve this discontinuity, higher merging scales have been applied to the calculation. Application of a merging scale of 70 GeV ($\log(70)=1.85$) does not lead to discontinuities  as with the 23 GeV merging scale (Fig. \ref{fig:etclus70}). 

By applying a larger merging scale to events with high di-lepton masses, the low multiplicity samples contribute more at higher scales. For example, $d_{01}$ is smoother when the $2\rightarrow 2$ sample contributes largely up to 70 GeV instead of falling down rapidly at scales larger than 23 GeV.

%\begin{figure}
%\begin{center}
%\includegraphics[width=0.45\textwidth]{DJR01_mmll600}
%\includegraphics[width=0.45\textwidth]{DJR01_mmll800}
%\caption{The $d_{01}$ distributions (scale at which a Z+0 jet configuration resolves in a Z+1 jet configuration) for a minimal DY mass of 600 GeV (left plot) and 800 GeV (right plot) with different merging scales. The bottom plots show the ratio with respect to the curve that has $\mu_m = 23$ GeV.}
%\label{fig:DJRs}
%\end{center}
%\end{figure}

\newpage

\section{Conclusion} \label{sec:conclusion}
%In this article I have studied DJRs in extreme kinematic situations of high mass DY production accompanied by QCD jets. I applied the TMD merging approach to perform these studies. 

Transverse momentum recoils in the initial-state shower, 
taken into account via TMD distributions, affect the 
theoretical systematics of predictions 
for jet final 
states \cite{Dooling:2012uw,Hautmann:2012dw,Hautmann:2013fla}. 
Focusing on the 
case of DJR variables in DY + jets production, in 
this article I have investigated, within a PB TMD framework 
with multi-jet merging, the dependence of predictions	 
on the merging scale as a function of the DY mass scale. 
I have shown in particular that the results support the 
possibility of a merging scale increasing with mass. 

It looks to be of interest to explore this possibility further, 
also in the context of current developments on TMD 
parton branching, such as the ongoing determination \cite{Barzani:2022msy} of 
PB TMD distributions via 
xFitter~\cite{xFitterDevelopersTeam:2022koz,Alekhin:2014irh}
including effects of dynamical 
soft-gluon resolution scales~\cite{Hautmann:2019biw} and 
angular ordering~\cite{Hautmann:2017fcj,Catani:1990rr,Marchesini:1987cf}, 
and the extension of the branching evolution to incorporate   
TMD splitting functions~\cite{Hautmann:2022xuc,Catani:1994sq}. Besides, it will be interesting to compare the behavior 
of the TMD merging algorithm with that of collinear merging algorithms 
when the DY mass scale is varied.

%Events of these kinematics are rare if no restrictions are set on the hard scale. 
%Conventional multi-jet merging approaches use a fixed merging scale, independent of the mass of the vector boson. The parton shower contributions are then very much limited and matrix elements description need to fill phase space regions that are effected by soft and collinear parton emissions. It is shown that the use of a low merging scale is not preferable in these type of events, since the DJR becomes discontinuous. The use of a merging scale that is larger improves the continuity of the DJR and therefore the quality of the merging.

%From this result, I conclude that the merging scale depends on the mass of the di-lepton pair. Further investigations are necessary to identify the exact relation between $\mu_m$ and $M_{ll}$ and implementation in the TMD merging algorithm. 

\section*{Acknowledgements}
I thank Nestor Perez Armesto and the other organizers of the DIS-2022 workshop for the possibility to present these results at this interesting conference. Many thanks to Francesco Hautmann and Armando Bermudez Martinez for collaboration and interesting discussions.

\bibliographystyle{mybibstyle} 

\bibliography{DIS22Ref}

\end{document}